\documentclass[fleqn,10pt]{olplainarticle}
\usepackage{url}
\usepackage{pdflscape}
\usepackage{adjustbox}
\usepackage{colortbl}
\usepackage{booktabs}

\title{Comparative analysis of Mixed-Data Sampling (MIDAS) model compared to Lag-Llama model for inflation nowcasting}

\author[1]{Adam Bahelka}
\author[2]{Harmen de Weerd}
\affil[1]{a.bahelka@student.rug.nl}
\affil[2]{harmen.de.weerd@rug.nl}


\begin{abstract}
Inflation is one of the most important economic indicators closely watched by both public institutions and private agents. This study compares the performance of a traditional econometric model, Mixed Data Sampling regression, with one of the newest developments from the field of Artificial Intelligence, a foundational time series forecasting model based on a Long short-term memory neural network called Lag-Llama, in their ability to nowcast the Harmonized Index of Consumer Prices in the Euro area. Two models were compared and assessed whether the Lag-Llama can outperform the MIDAS regression, ensuring that the MIDAS regression is evaluated under the best-case scenario using a dataset spanning from 2010 to 2022. The following metrics were used to evaluate the models: Mean Absolute Error (MAE), Mean Absolute Percentage Error (MAPE), Mean Squared Error (MSE), correlation with the target, R-squared and adjusted R-squared. The results show better performance of the pre-trained Lag-Llama across all metrics.
\end{abstract}

\begin{document}

\flushbottom
\thispagestyle{empty}
\maketitle
\UseRawInputEncoding

\section{Introduction}

Inflation nowcasting, the practice of predicting the current rate of inflation, is crucial for both policymakers and market participants \citep{monteforte-2012}. However, a significant challenge in inflation nowcasting is the issue of mixed-frequency data. Inflation is typically measured on a monthly basis, but many relevant indicators, such as financial market data or commodity prices, are available on a daily basis. Traditional econometric models, such as Autoregressive Integrated Moving Average (ARIMA) or Vector Autoregression (VAR), are not equipped to handle data of varying frequencies. A common workaround is to aggregate high-frequency data to match the frequency of the target variable, but this approach can lead to the loss of valuable information.

The Mixed Data Sampling (MIDAS) model, developed by \cite{ghysels-2004, ghysels-2005, ghysels-2007}, offers a solution by allowing the use of high-frequency data without the need for aggregation \citep{knotek2017nowcasting, kronenberg2023nowcasting}. This model has been widely adopted by public institutions and private entities due to its ability to enhance forecasting accuracy by incorporating real-time data that reflects the latest market conditions or economic indicators \citep{knotek2017nowcasting, kronenberg2023nowcasting}. Recently, advancements in machine learning have introduced new models for time series forecasting. One such model is Lag-Llama, a foundational time series forecasting model based on Long Short-Term Memory (LSTM) neural networks. Developed by \cite{rasul2024lagllama}, Lag-Llama is trained on a large number of time-series datasets and can perform zero-shot forecasting, making accurate predictions on datasets it has never been explicitly trained on. However, Lag-Llama does not support mixed-frequency data, which can limit its effectiveness in applications where the target variable and predictors are measured at different frequencies. In this study, we compare the performance of the MIDAS model and the Lag-Llama model in nowcasting the Harmonized Index of Consumer Prices (HICP) in the Euro area. Additionally, we compare these models to a simple benchmark: an autoregressive model that uses one previous value, known as AR(1). By evaluating these models performance, with the MIDAS regression under optimal conditions, our study aims to introduce foundational time-series forecasting models into the field of econometrics.

\subsection{Problem Description}

Inflation is a crucial economic metric for both institutional and private operators. For institutions, timely economic data and projections are essential for macroeconomic policy \citep{monteforte-2012}. Central banks, for instance, increasingly base their monetary policy on variations of the Taylor rule \citep{castro2011can}, which describes the relationship between inflation and other economic variables, namely interest rates and the output gap \citep{taylor1993discretion}. \cite{taylor1993discretion} argues that central banks can manage inflation through changes in interest rates. As for the private agents, reliable inflation data is important for adjusting investment strategies \citep{monteforte-2012} and for successful price setting \citep{ang2007macro}. Typically, as inflation is measured on a monthly basis, a common approach to forecasting inflation is to construct models with monthly frequency data. However, these models lack very recent information that can be found in daily data \citep{monteforte-2012}. Furthermore, data on inflation is often published with a time lag \citep{monteforte-2012}. For example, the U.S. Bureau of Labor Statistics publishes its consumer price index for a given month only in the middle of the following month\footnote{These dates and times can be found at: \url{https://www.bls.gov/schedule/news_release/cpi.htm}}. Its EU counterpart Eurostat follows a very similar timeline.\footnote{Dates for releasing the HICP can be found here: \url{https://ec.europa.eu/eurostat/news/euro-indicators/release-calendar?start=1714514400000&type=listMonth&indicator=Inflation}} Due to this time lag, there is a space for inflation nowcasting, which involves forecasting inflation for the near future, present, or recent past.

\subsection{Background}

Over the years, research on improving inflation predictions has advanced. Numerous studies have examined different models suitable for this task. In this research, we focus on MIDAS regression and the Lag-Llama LSTM model.

\subsubsection*{MIDAS regression background}

Several econometric models are suitable for time series forecasting, such as Autoregressive Integrated Moving Average (ARIMA) developed by \cite{BoxJenkins1970} and Vector Autoregression (VAR) developed by \cite{sims1980macroeconomics}. However, none of these can handle mixed-frequency data. When it comes to nowcasting, this leads to a loss of valuable information present in the higher frequency data. A common approach is data aggregation, but this approach results in the loss of some information \citep{ghysels-2004}. Several models in econometrics can combine mixed-frequency data. We focus on a model known as Mixed Data Sampling (MIDAS) regression, developed by \cite{ghysels-2004, ghysels-2005, ghysels-2007}, to integrate data of different frequencies. The primary advantage of MIDAS regression is that it allows for the use of higher-frequency data such as daily or weekly measurements without the need to aggregate this information to the frequency of the target variable, typically observed at lower frequencies like monthly or quarterly intervals.\\

One of the primary advantages of MIDAS regressions is their ability to enhance forecasting models by incsorporating real-time data that reflects the latest market conditions or economic indicators \citep{ghysels-2007}. This inclusion not only improves the timeliness of the forecasts but can also increase their accuracy, which is crucial for both macroeconomic planning and financial investment strategies. In practice, MIDAS models have been implemented by several research institutes and public institutions, including the Federal Reserve Bank of St. Louis \citep{knotek2017nowcasting} and the ETH Nowcasting Lab \citep{kronenberg2023nowcasting}. They use these models to enhance the precision of economic forecasts that inform monetary policy decisions. The ability of MIDAS models to include high-frequency financial indicators, such as stock prices or interest rate spreads, provides a more nuanced view of economic trends, allowing policymakers to react more swiftly and appropriately to economic shifts \citep{ghysels-2007}. \\

In conclusion, the MIDAS regression represents a traditional econometric model specifically designed to incorporate mixed-frequency data, thereby improving prediction accuracy in tasks where this is necessary. As economic and financial data continue to grow in volume and velocity, the relevance and utility of MIDAS models may increase, given their ability to adjust forecasts in real-time.

\subsubsection*{Machine Learning Background}

In the past three decades, the rapid increase in computational power and resources in the field of machine learning has constantly developed new ways of predicting tasks, including time-series predictions. One of the first significant developments was the introduction of Long Short-term Memory neural networks (LSTM) by \cite{6795963}. Over the years, improvements in architecture, such as the introduction of the attention mechanism \citep{bahdanau2016neural} and transformers \citep{vaswani-2017}, have improved their efficiency and capabilities. Although neural network models are capable of achieving great results, they are typically trained for specialized tasks. These specialized models have already been the subject of studies and comparisons with different econometric models, such as ARIMA or VAR \citep{siami2018comparison, parmezan2019evaluation}. These studies have found that machine learning models can often outperform traditional models like ARIMA and VAR. \\

There are a few challenges in the deployment of machine learning models. First, substantial computational power and data are necessary to train a machine learning model, resources often unavailable to many institutions and private entities. In recent years, machine learning has experienced a paradigm shift due to the development of foundation models \citep{bommasani2022opportunities}. These are large-scale general-purpose networks trained on large datasets and designed to be adaptable to a wide range of tasks without requiring task-specific training. This approach gained significant traction following the introduction of large-scale language models such as ChatGPT models \citep{openai2024gpt4} or the open-source model LLaMA \citep{touvron2023llama}. The architectural principles of LLaMA have also been applied in the development of Lag-Llama, one of the pioneering foundational models for time series forecasting. Lag-Llama is distinguished by its ability to perform zero-shot forecasting, enabling it to make accurate predictions on datasets it has never been explicitly trained on \citep{rasul2024lagllama}. This pre-trained model combines the robustness of specific models with lower resource demands on computing power or data.\\

\subsection{Research Question}

The aim of the study is to compare the MIDAS model with a pre-trained Lag-Llama model in their ability to capture changes in monthly inflation based on various daily data. The reason for choosing these two models is that the MIDAS model symbolizes the state-of-the-art model often used by public institutions and private organizations. Lag-Llama, on the other hand, is the result of recent developments in the machine learning field and its relative ease of use has the potential to spread among public institutions and private agents. The aim of the study is therefore to determine which of these two models works better in the task of predicting inflation. This gives us the following research question: 

\textbf{Can a pre-trained neural network Lag-Llama model outperform the MIDAS regression across all the performance metrics in inflation nowcasting?}

\section{Methods}
In this section, we will describe the data used and provide a detailed theory and implementation of the two previously mentioned models.

\subsection{Data}
\subsubsection{Data choice and description}

For this study, a dataset was created based on previous research on MIDAS modelling, spanning from June 2010 to June 2022. From the literature, specifically daily data that are available in real time were chosen. The advantage of this is that it makes it suitable for nowcasting by being able to adjust the forecast with the introduction of new data every day. Therefore, predictions can dynamically change as soon as the new data are published. Our dependent variable is the Harmonized Index of Consumer Prices (HICP)\footnote{HICP retrieved from \url{https://ec.europa.eu/eurostat/web/hicp/overview}} with a monthly frequency. This index captures the measurement of inflation in the Euro area and ensures the same method is used for calculating inflation in all Eurozone countries. HICP represents the average change in household spending for various goods and services \citep{ECB_HICP}.\\
Our independent variables are daily prices of Brent crude oil\footnote{Brent crude oil prices retrieved from \url{https://fred.stlouisfed.org/series/DCOILBRENTEU}}. This represents potential supply shocks in the economy as oil prices have an effect on inflation \citep{blanchard_macroeconomics_2021}, and this data was also used by the Federal Reserve Bank of St. Louis \citep{knotek2017nowcasting}. Furthermore, we are using daily data on the interest rate spread calculated as the difference between the price of the 10-year German Bund\footnote{Prices of 10-year German Bunds retrieved from: \url{https://www.investing.com/rates-bonds/germany-10-year-bond-yield-historical-data}} and the 3-month German interbank rate\footnote{90-day German interbank rate retrieved from: \url{https://fred.stlouisfed.org/series/IR3TIB01DEM156N}}, short-term interest rates represented by the 3-month Libor\footnote{LIBOR retrieved from: \url{http://iborate.com/eur-libor/}} and long-term interest rates represented by the price of the 10-year German Bund. All these data points are supposed to reveal the latest market and inflation expectations and were used by \citep{monteforte-2012}. Moreover, we are using USD/EUR\footnote{USD/EUR exchange rate data retrieved from: \url{https://www.investing.com/currencies/usd-eur-historical-data}} exchange rates as well as CNY/EUR\footnote{CNY/EUR exchange rate data retrieved from: \url{https://www.investing.com/currencies/cny-eur-historical-data}} exchange rates as they can capture the changes in import and export prices, as used by \cite{monteforte-2012}. Lastly, the Eurostoxx50\footnote{Eurostoxx50 index data retrieved from \url{Eurostoxx50}} index was chosen, as stock market indices contain information about inflation expectations \citep{stock1999forecasting}. In Table \ref{table:stats_summary} in the Appendix, we can see the descriptive statistics of our variables. The statistics include measures such as the count, mean, standard deviation, minimum, maximum, and various percentiles. \\

\subsubsection{Data pre-processing and post-processing}

\begin{figure}[h]
    \centering 
    \includegraphics[width=0.8\linewidth]{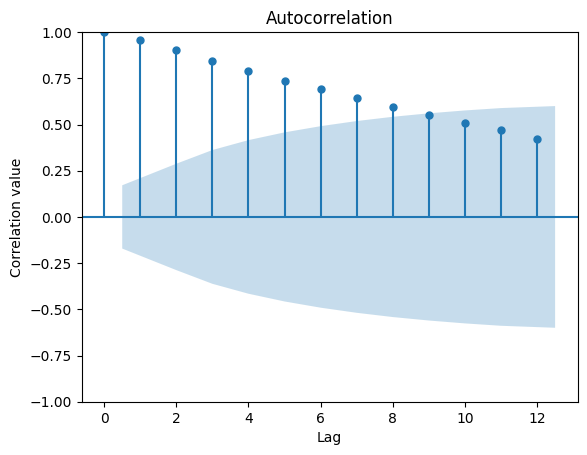} 
    \caption{Autocorrelation of HICP, measuring the correlation coefficients between the time series and HICP's past values}
    \label{fig:autocorrelation} 
\end{figure}

First, we will focus on our dependent variable. A common way of analyzing a time series is autocorrelation. Autocorrelation measures the correlation between a time series' own past values, indicating how past data points influence future data points. It helps identify patterns such as trends or seasonal effects within the data. The blue shaded area in the autocorrelation plot represents the 5\% confidence interval for the autocorrelation coefficients at each lag. Our autocorrelation results are shown in Figure \ref{fig:autocorrelation}, which shows significant positive correlation values for the initial 8 lags, gradually declining towards zero. This indicates that the past three quarters of a year might have an influence on our forecast. \\

\begin{figure}[h]
    \centering 
    \includegraphics[width=0.8\linewidth]{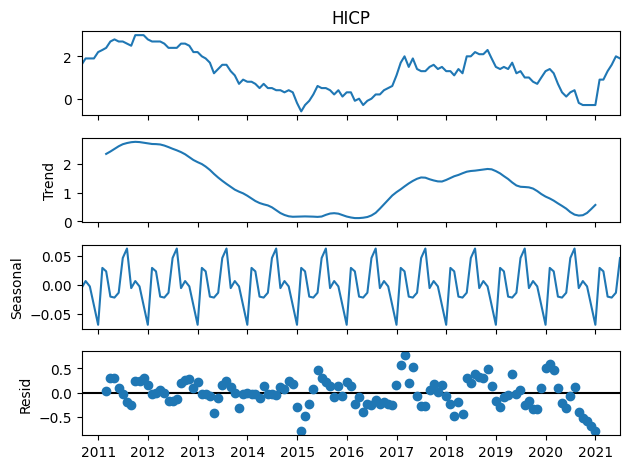} 
    \caption{Deseasoning of HICP} 
    \label{fig:deseasoning} 
\end{figure}

A common practice in time series analysis is seasonal decomposition. Seasonal decomposition is a method used to analyze time series data by breaking it down into three main components: trend, seasonality, and residual (noise). The trend component represents the long-term progression of the series, showing the general direction in which the data is moving. This is done by applying a centered moving average to the time series with a window of 12 months. This helps to smooth out short-term fluctuations and highlight longer-term trends. The seasonality component captures the repeating yearly short-term cycle in the data. This is done by averaging the data points at corresponding yearly periods across multiple cycles. Lastly, the residual component represents the irregular fluctuations that are not explained by the trend or seasonality. The results of our decomposition are shown in Figure \ref{fig:deseasoning}. The top plot shows the original HICP series, highlighting the overall trends and fluctuations from 2010 to 2021. The second plot depicts the trend component, which captures the long-term movement in the data, showing a general decline until around 2016, followed by a slight increase and subsequent stabilization. The third plot illustrates the seasonal component, which represents the regular, repeating patterns observed within each year. Although there was a weak seasonality found between $0.05 -0.05$ of inflation change, the data from the European Central Bank are already seasonally adjusted. Finally, the residual component in the bottom plot shows the irregular fluctuations or noise remaining after removing both the trend and seasonal effects. \\

Although we found a trend that could represent a form of economic cycles in the economy, this trend cannot be fitted easily into a simple linear or polynomial equation. Therefore, after separating the trend from the target data, another forecasting model would be necessary for the trend of our data, which likely would not increase the performance of our predictions. For this reason, we have decided not to de-trend our data as it most likely would not improve our forecasts.\\

For the independent variables, the daily values of fractional changes in the past 7 days were calculated. For example, the calculated data for the variable $OIL$ is shown in Equation \ref{eq:fractional_change_oil}. This normalizes the data as different variables are using different scales. \\

\begin{equation}
\Delta \text{OIL}_t = \frac{\text{OIL}_t - \text{OIL}_{t-7}}{\text{OIL}_{t-7}}
\label{eq:fractional_change_oil}
\end{equation}

For the Lag-Llama model, there was additional pre-processing of the data. This consisted of downsampling the target variable to a daily frequency by filling in the values for the respective month. This was necessary to make it suitable for the Lag-Llama model as the model doesn't support mixed frequency input data. \\

\subsection{Multicollinearity and Variance Inflation Factor (VIF)}
In the context of econometric modeling, multicollinearity refers to the scenario where predictor variables are highly correlated, which can undermine the statistical significance of individual predictors. To assess multicollinearity in our dataset, we computed the Variance Inflation Factor (VIF) for each predictor variable. The VIF quantifies how much the variance of a regression coefficient is inflated due to multicollinearity in the model. A VIF value above 5 generally indicates high multicollinearity, which could be problematic for model estimation \cite{kim2019multicollinearity}.\\

In our analysis, the VIF values for each predictor converge to values close to 1, except for the USD/EUR exchange rates and CNY/EUR exchange rates with VIF values of 1.4 and 1.37 respectively. This suggests that multicollinearity is not expected to be an issue in our dataset. Table \ref{tab:VIF_factors} presents the VIF values for all predictors used in our models.\\

\begin{table}[h]
\centering
\caption{VIF Factors for Predictor Variables}
\label{tab:VIF_factors}
\begin{tabular}{lc}
\toprule
\textbf{Features} & \textbf{VIF Factor} \\
\midrule
\textbf{const} & 1.053 \\
\textbf{OIL} & 1.104 \\
\textbf{LIBOR\_3M} & 1.018 \\
\textbf{INTEREST\_SPREAD} & 1.001 \\
\textbf{USD/EUR exchange rate} & 1.400 \\
\textbf{Eurostoxx50\_index} & 1.089 \\
\textbf{CNY/EUR exchange rate} & 1.366 \\
\textbf{German\_bund} & 1.001 \\
\bottomrule
\end{tabular}
\end{table}

\begin{figure}[h]
    \centering 
    \includegraphics[width=1\linewidth]{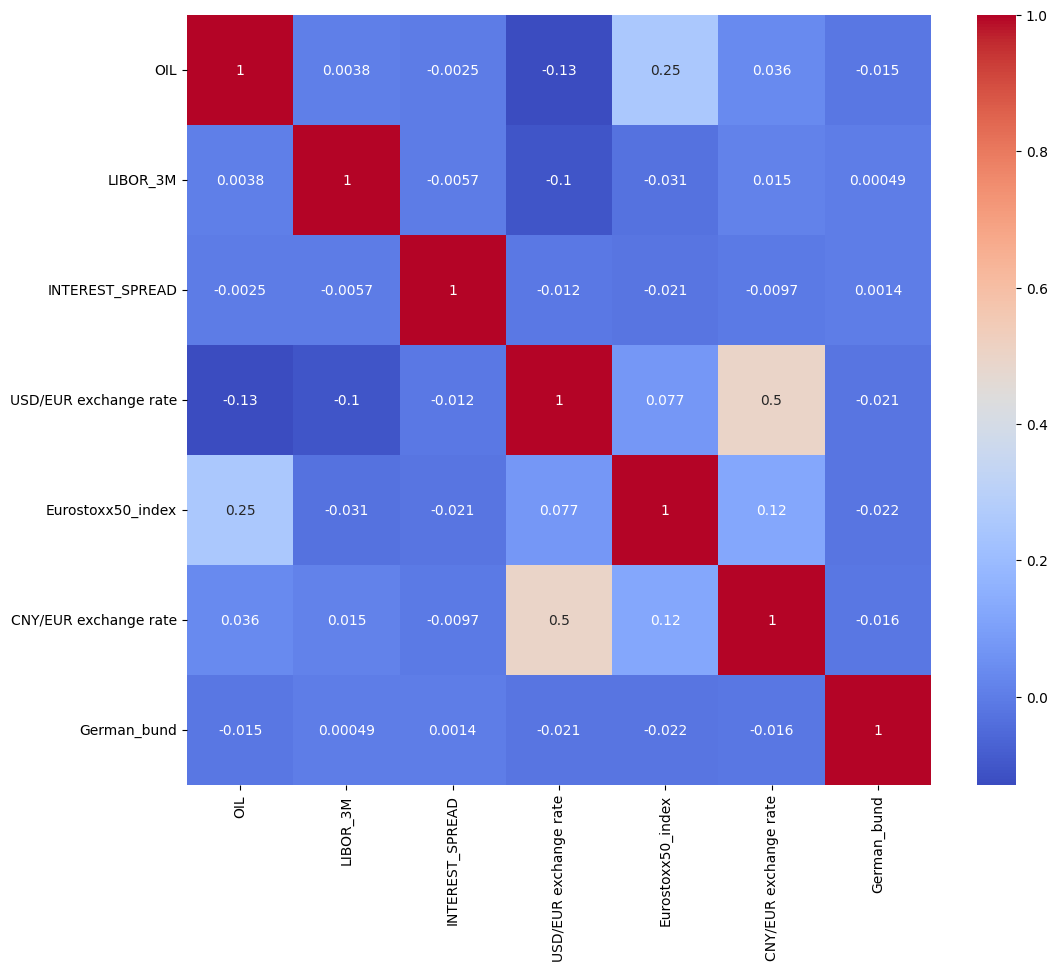} 
    \caption{Correlation matrix of independent variables} 
    \label{fig:corr_matrix} 
\end{figure}

To further understand the relationships between the predictors, a correlation matrix was computed, as shown in Figure \ref{fig:corr_matrix}. This matrix helps to identify the degree of the linear relationship between each pair of variables. The correlation matrix reveals that most predictors have low to moderate correlations with each other, with the highest correlation being 0.5 between the USD/EUR exchange rate and the CNY/EUR exchange rate. This further confirms that multicollinearity is not expected to be an issue in our dataset as multicollinearity typically starts to be present at correlation coefficients of around 0.8 among the independent variables \citep{shrestha2020detecting}.\\

\subsection{Theory behind MIDAS}
MIDAS is a  type of regression that is able to accommodate for predicting the target sampled with a lower frequency while its features can be sampled at a higher frequency. First, we will introduce the theory behind the model, followed by a polynomial specification for the model. \\
\subsubsection*{Theory}
Based on the theory of distributed lag models, \cite{ghysels-2004} has introduced a slightly different approach in order to accommodate data sampled at different frequencies. A simple MIDAS regression with one endogenous variable $y_t$ and one exogenous variable $X$ is shown in Equation \ref{eq:basic_midas}. Suppose that $y$ is sampled at some fixed frequency, in our case monthly. Now let $X^{(m)}$ be sampled $m$ times faster. In our case, this would be daily, which gives us approximately $m=30$. With this notation, the simple MIDAS regression model would have the following form:\\
\begin{equation}
y_t = \beta_0 + \beta_1 B(L^{1/m}; \theta)X_t^{(m)} + \epsilon_t^{(m)}
\label{eq:basic_midas}
\end{equation}
\\
where: 
\begin{equation}
B(L^{1/m}; \theta)= \sum_{j=0}^{J} B(j; \theta)L^{j/m}
\label{eq:basic_midas_1}
\end{equation}
 Equation \ref{eq:basic_midas_1} shows a polynomial of length $J$ where $B(j; \theta)$ is the weighting function, $\theta$ is the vector of hyperparameters defining the weighting function. This describes how each lagged value of $X^{(m)}_{t}$ is weighted in the regression. The lagged polynomial notation $(L^{j/m})$ shows the period of our endogenous variable as shown in Equation \ref{eq:basic_midas_2}.
 \begin{equation}
L^{j/m}X_t^{(m)} = X^{(m)}_{t-j/m}
\label{eq:basic_midas_2}
 \end{equation}\\
 
To put it in another way, the $L^{j/m}$ operator produces the value of $X^{(m)}_t$ lagged by $j/m$ periods. In our example, this would mean that the above equation is a prediction of monthly inflation ($y_t$) on the basis of daily data ($X^{(m)}_t$) using up to $J$ daily lags (our context horizon). \\

Based on this theory, we can reach an immediate conclusion. The MIDAS regression model is optimized over the vector of hyperparameters in $\theta$ which results in $J$ parameters to estimate for every exogenous variable, namely $B(L^{1/m}; \theta)$. When adding more exogenous variables, we might face a problem of parameter proliferation, or as it is called in the machine learning community, the "curse of dimensionality". This is the trade-off when constructing a MIDAS regression. There is a benefit in exploiting data of higher frequencies compared to aggregating all the data to the same low frequency. However, when adding too many variables with high frequency, there is a risk of overfitting the model. \\

\subsubsection*{Polynomial specification}
\begin{figure}[h]
    \centering 
    \includegraphics[width=0.8\linewidth]{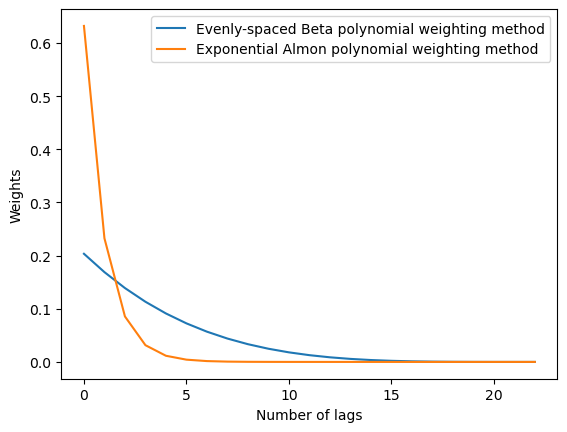} 
    \caption{Almon and Beta functions with 22 lags} 
    \label{fig:parameter_functions} 
\end{figure}

One of the important parts of the MIDAS architecture is the choice of the parametrization of $B(j; \theta)$. This is essentially the function that will describe how different lags will have different weights in our regression. One popular function is the exponential Almon lag, which originated from the literature about distributed lag models \citep{almon1965distributed}. Another popular polynomial is the Beta lag introduced by \cite{ghysels-2005}. Figure~\ref{fig:parameter_functions} shows the difference between Almon and Beta polynomials. \cite{ghysels-2007} argues that variables with Beta lag have better statistical significance, which is also why we used them in our research. \\

\subsection{Lag-Llama}
Following the new paradigm shift created by foundation models trained on large datasets \citep{bommasani2022opportunities}, Lag-Llama is an attempt to bring this success to the field of time-series forecasting. Lag-Llama is an LSTM neural network with a decoder-only transformer-based architecture that was inspired by Llama \citep{touvron2023llama}. \cite{rasul2024lagllama} took the well-known architecture from the Llama model and changed tokenization and prediction in order to accommodate for time series forecasting. Instead of tokenizing text into individual words, as is standard practice in LLMs, tokenization constructs lagged features from the prior values of the time series also with a time series indices (this can be second, hour, day, month...). Besides the lagged features, \cite{rasul2024lagllama} added date-time features to provide additional information such as the second of a minute, hour of the day, and up to the quarter of the year. This inclusion allows the model to implicitly understand the frequency of the series and provide additional temporal context that may influence predictions.\\

When it comes to the last layer of the architecture that is making predictions, Lag-Llama utilizes a so-called distribution head, which projects the features to the parameters of a probability distribution.  A Student's t-distribution is used \citep{student-1908} in the distribution head that outputs the degrees of freedom, mean, and scale of the distribution.  This is a difference compared to MIDAS regression that isn't able to perform probabilistic forecasting as it produces point estimates based on ordinary least squares (OLS) method of lagged values. \\

\subsection{Implementation}
The Python code with implementation of both models can be found on GitHub.\footnote{Link to the GitHub repository: \url{https://github.com/AdamBahelka/MIDAS_LagLama_inflation_nowcasting.git}}
\subsubsection*{Implementation of the MIDAS model}
Python was used for the implementation of both models. The implementation of the MIDAS model was made possible by the work by Yoseph Zuskin\footnote{Class for MIDAS model can be found at \url{https://github.com/Yoseph-Zuskin/midaspy?tab=readme-ov-file}}. For the model, the horizon was chosen at 22 lags for the exogenous variable and 1 lag for the endogenous variable, as the same values were used used by \cite{ghysels-2007}. 
\subsubsection*{Implementation of Lag-Llama}
The implementation of the Lag-Llama model was made according to \cite{rasul2024lagllama}. This model can be found on Hugging Face\footnote{You can access the model on Hugging Face at \url{https://huggingface.co/time-series-foundation-models/Lag-Llama}} and on GitHub.\footnote{You can access the model on GitHub at \url{https://github.com/time-series-foundation-models/lag-llama}} For the model, the context length that demonstrates the best performance was 2400 points of daily data, around seven and a half years. A rolling window was constructed to predict the next month on the basis of previous inflation data and the chosen daily features for the above mentioned seven and a half year period. \\

\subsection{Performance measures}
In order to compare the two models, we have looked at multiple measures. Our null hypothesis was that there is no difference between the performance of the two models. In order to answer our research question, we have evaluated whether the Lag-Llama model can outperform MIDAS regression. As mentioned, MIDAS is given the best case scenario, evaluating it on in-sample predictions and on the specialized task of mixed frequency data that Lag-Llama is not optimized for. We have defined Lag-Llama as outperforming the MIDAS model only if all the below mentioned performance metrics were lower than 95\% (or higher than 105\% where it is appropriate) compared to the MIDAS regression. \\
\subsubsection*{Mean absolute error}
The mean absolute error (MAE) is one of the simplest ways of evaluating a model. The equation calculates the average of the difference between the predicted value and the target value. The formula for calculating MAE is: 
\begin{equation}
\text{MAE} = \frac{1}{n} \sum_{i=1}^n \left| y_i - \hat{y}_i \right|
\label{eq:mae}
\end{equation}
where $y_i$ is the true value of inflation and $\hat{y}_i$ is the prediction of our model. In our case, $n=24$ as we are evaluating the model on two years of data. 
\subsubsection*{Mean absolute percentage error}
Mean absolute percentage error (MAPE) shows a similar statistic as Equation \ref{eq:mae}, but as a percentage difference between the target value and predicted value. The equation is as follows: 
\begin{equation}
\text{MAPE} = \frac{100}{n} \sum_{i=1}^n \left| \frac{y_i - \hat{y}_i}{y_i} \right|
\label{eq:mape}
\end{equation}
\subsubsection*{Mean squared error}
The mean squared error (MSE) is a measure that evaluates the accuracy of the predictions. The equation is used to calculate the average of the difference between the predicted value and the target value. Because of its squared term, this method imposes proportionally greater penalty the further the prediction is from the target. The formula for calculating MSE is:\\
\begin{equation}
\text{MSE} = \frac{1}{n} \sum_{i=1}^n (y_i - \hat{y}_i)^2
\label{eq:mse}
\end{equation}

\subsubsection*{Correlation}
Another measure suitable for the evaluation of the quality of predictions is the correlation. We will use the formula for the Pearson correlation coefficient between the predicted time-series and the actual time-series. The formula for calculating the correlation is:
\begin{equation}
r = \frac{\sum_{i=1}^n (y_i - \bar{y})(\hat{y}_i - \bar{\hat{y}})}{\sqrt{\sum_{i=1}^n (y_i - \bar{y})^2}\sqrt{\sum_{i=1}^n (\hat{y}_i - \bar{\hat{y}})^2}}
\label{eq:pearson}
\end{equation}
where \( y_i \) represents the actual values of the data, \( \hat{y}_i \) represents our predicted values, \( \bar{y} \) is the mean of the actual values, and \( \bar{\hat{y}} \) is the mean of the predicted values. The numerator of the formula, \( \sum_{i=1}^n (y_i - \bar{y})(\hat{y}_i - \bar{\hat{y}}) \), gives the covariance between the actual and predicted values, which measures how changes in the actual values are associated with changes in the predicted values. The denominator normalizes this covariance by the product of the standard deviations of both the actual and predicted values, ensuring the result is scaled to a value between $-1$ and 1.\\

\subsubsection*{Benchmark model}
The benchmark model used in this study represents a simple and effective model, often referred to as the naive model. It relies on the assumption that the best predictor of the future value of a time series is its most recent observed value. In the context of this study, the benchmark model is a linear regression that predicts the monthly inflation rate by using the inflation rate of the previous month. This model serves as a comparison and is expected to be outperformed by Lag-Llama and MIDAS regression. Equation \ref{eq:benchmark} shows the simple structure of our benchmark model.

\begin{equation}
\hat{y}_{t} = \beta y_{t-1}
\label{eq:benchmark}
\end{equation}
where \( \hat{y}_{t} \) is the predicted inflation rate for the current month, \( y_{t-1} \) is the observed inflation rate of the previous month, and $\beta$ is the OLS coefficient of our regression. This benchmark model's performance was be evaluated using the same metrics as the other models to provide a clear comparison.\\

\section{Results}
In this section, we present the findings of our comparative study between the MIDAS and Lag-Llama model on the task of nowcasting Eurozone HICP. Initially, we assess the significance of individual predictors within the MIDAS model, as detailed in the regression results table, which clarifies how each factor contributes to the model's predictive capabilities. Subsequently, we have compared the predictive accuracy of MIDAS and Lag-Llama, emphasizing their performance over a critical two-year evaluation period. 
\subsection{Results of the MIDAS Model}
In this section, we present the findings from the MIDAS model, focusing on its predictive accuracy for nowcasting the HICP in the Euro area. The MIDAS model's predictions were generated under optimal conditions to ensure a fair and accurate comparison with the Lag-Llama model. 

\subsubsection{Significance of Predictors}
Table~\ref{tab:regression_results} shows the estimated coefficients obtained from the regression model. Each row in the table corresponds to a different predictor variable, with the following columns providing detailed statistical metrics:

\begin{itemize}
    \item \textbf{coef}: The name of the predictor variable.
    \item \textbf{Estimate}: The estimated effect of the predictor on the dependent variable, holding all other predictors constant.
    \item \textbf{std err}: The standard error of the estimate\textbf{,} which measures the average amount that the estimate will differ from the actual value.
    \item \textbf{t}: The ratio of the estimate to its standard error. Higher absolute values of the t-value indicate greater statistical significance.
    \item \textbf{P$>|t|$}: The p-value associated with the hypothesis test that the coefficient is equal to zero (no effect). A small p-value (typically less than 0.05) indicates strong evidence against the null hypothesis, thus suggesting a significant effect.
    \item \textbf{[0.025 0.975]}: The 95\% confidence interval for the coefficient estimate. If this interval does not include zero, the effect is considered statistically significant at the 5\% level.
\end{itemize}

\begin{table}[htbp]
\centering
\caption{Regression Results of the MIDAS model}
\label{tab:regression_results}
\begin{tabular}{lcccccc}
\toprule
 & \textbf{coef} & \textbf{std err} & \textbf{t} & \textbf{P$>|t|$} & \textbf{0.025} & \textbf{0.975} \\
\midrule
\textbf{Constant} & 0.092 & 0.040 & 2.320 & 0.022 & 0.014 & 0.171 \\
\textbf{OIL} & 0.821 & 0.273 & 3.003 & 0.003 & 0.280 & 1.362 \\
\textbf{LIBOR\_3M} & -0.005 & 0.104 & -0.046 & 0.963 & -0.210 & 0.201 \\
\textbf{INTEREST\_SPREAD} & 0.108 & 0.050 & 2.190 & 0.030 & 0.010 & 0.206 \\
\textbf{USD/EUR exchange rate} & 2.224 & 1.293 & 1.720 & 0.088 & -0.334 & 4.781 \\
\textbf{Eurostoxx50\_index} & -2.087 & 0.916 & -2.279 & 0.024 & -3.899 & -0.275 \\
\textbf{EUR/CNY exchange rate} & -3.621 & 1.809 & -2.002 & 0.047 & -7.200 & -0.042 \\
\textbf{German\_bund} & -0.008 & 0.012 & -0.717 & 0.475 & -0.031 & 0.015 \\
\textbf{HICP t-1} & 0.935 & 0.025 & 37.147 & 0.000 & 0.885 & 0.985 \\
\bottomrule
\end{tabular}
\end{table}
The results indicate significant relationships between several predictors and the dependent variable. For instance, the coefficient for \textit{OIL} is positive and significant, indicating a strong relationship with the dependent variable (\(\textit{coef} = 0.821\), \(P = 0.003\)). Contrary to previous findings \citep{monteforte-2012}, we find no significant effect of short-term interest rates, represented by 3-month euro LIBOR rates, on inflation (\(\textit{coef} = -0.005\), \(P = 0.963\)). The same can be said about the relation between long-term interest rates, represented by the prices of German Bunds, and inflation (\(\textit{coef} = -0.008\), \(P = 0.475\)). Another significant variable is the relatively large negative coefficient for \textit{Eurostoxx50\_index} (\(\textit{coef} = -2.087\), \(P = 0.024\)) and \textit{EUR/CNY exchange rate} (\(\textit{coef} = -3.621\), \(P = 0.047\)). The interest rate spread (\textit{INTEREST\_SPREAD}) is significant at the 5\% significance level (\(\textit{coef} = 0.108\), \(P = 0.030\)). The \textit{USD/EUR exchange rate} shows a positive yet marginally insignificant relationship (\(\textit{coef} = 2.224\), \(P = 0.088\)). As expected, the variable \textit{$HICP_{t-1}$}, which corresponds to the previous month's inflation, shows a strong positive correlation with the dependent variable (\(\textit{coef} = 0.935\), \(P < 0.001\)). 

\subsubsection{Evaluation Metrics}

\begin{figure}[h]
    \centering 
    \includegraphics[width=0.8\linewidth]{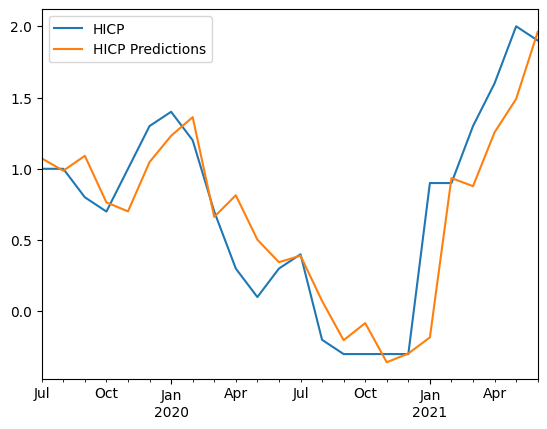} 
    \caption{Predictions of the MIDAS regression} 
    \label{fig:MIDAS_prediction} 
\end{figure}

The MIDAS model's predictions are compared against actual HICP values in Figure~\ref{fig:MIDAS_prediction}. The figure highlights the model's ability to follow the general trend of inflation but also reveals some deviations from the actual values. The performance metrics for the MIDAS model over the 24-month evaluation period and compared to the benchmark model are summarized in Table~\ref{tab:midas_vs_benchmark}:
\begin{itemize}
    \item \textbf{MAE}: The Mean Absolute Error for the MIDAS model is 0.23, indicating the average absolute deviation of predictions from actual values.
    \item \textbf{MAPE}: The Mean Absolute Percentage Error is 0.50, reflecting the average percentage error.
    \item \textbf{MSE}: The Mean Squared Error is 0.11, showing the average of the squared differences between predictions and actual values.
    \item \textbf{Correlation}: The Pearson correlation coefficient is 0.77, suggesting a strong linear relationship between predicted and actual values.
    \item \textbf{R-squared}: The R-squared value is 0.77, indicating that 77\% of the variance in the dependent variable is explained by the model.
    \item \textbf{Adjusted R-squared}: The Adjusted R-squared value is 0.72, which adjusts the R-squared value based on the number of predictors in the model.
\end{itemize}
Overall, the MIDAS model shows relatively good performance in inflation nowcasting, with all the metrics except the Correlation coefficient being lower than 95\% (or higher than 105\% where appropriate) compared to the metrics of the Benchmark model (these numbers are highlighted in bold in Table \ref{tab:midas_vs_benchmark}).

\begin{table}[htbp]
\centering
\caption{Comparison of MIDAS Regression to the Benchmark}
\label{tab:midas_vs_benchmark}
\begin{tabular}{lcc}
\toprule
\textbf{Metric} & \textbf{MIDAS} & \textbf{Benchmark} \\
\midrule
MAE & \textbf{0.23} & 0.26 \\
MAPE & \textbf{0.50} & 0.54 \\
MSE & \textbf{0.11} & 0.13 \\
Correlation & 0.88 & 0.86 \\
R-squared & \textbf{0.77} & 0.73 \\
Adjusted R-squared & \textbf{0.72} & 0.68 \\
\bottomrule
\end{tabular}
\end{table}

\subsection{Results of the Lag-Llama Model}

In this section, the performance of the Lag-Llama model, a neural network based on an LSTM architecture, in nowcasting the HICP for the Euro area is discussed. This model was evaluated over the same 24-month period from June 2019 to June 2021.\\

\subsubsection{Evaluation Metrics}
\begin{figure}[h]
    \centering 
    \includegraphics[width=0.8\linewidth]{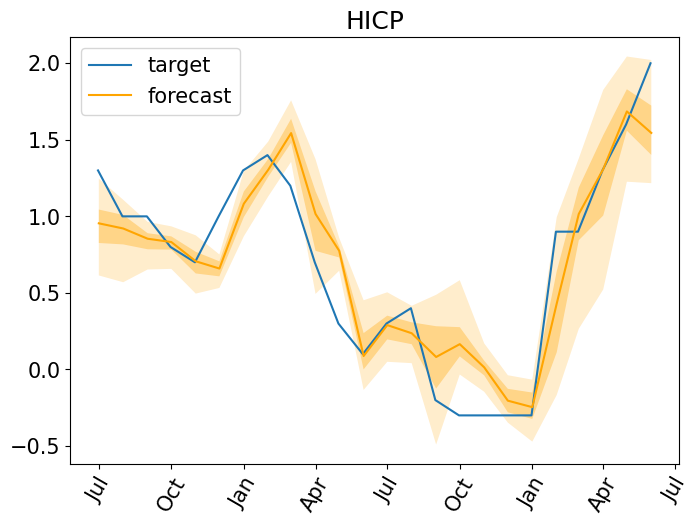} 
    \caption{Predictions of the Lag-Llama model} 
    \label{fig:Llama_prediction} 
\end{figure}

The Lag-Llama model's predictions are illustrated in Figure~\ref{fig:Llama_prediction}. This figure includes confidence intervals, providing a probabilistic view of the model's predictions. The visual comparison with actual HICP values suggests that the Lag-Llama model captures the overall inflation trend more closely than the MIDAS model. The performance metrics for the Lag-Llama model compared to the benchmark model are presented in Table \ref{tab:lag_llama_vs_benchmark}
\begin{itemize}
\item \textbf{MAE}: The Mean Absolute Error for Lag-Llama is 0.21, showing a smaller average deviation from actual values than the MIDAS model.
\item \textbf{MAPE}: The Mean Absolute Percentage Error is 0.40, lower than that of the MIDAS model, indicating better percentage accuracy.
\item \textbf{MSE}: The Mean Squared Error is 0.07, significantly lower than the MIDAS model, highlighting its effectiveness in minimizing large errors.
\item \textbf{Correlation}: The Pearson correlation coefficient is 0.92, higher than that of the MIDAS model, indicating a stronger alignment with actual inflation data.
\item \textbf{Adjusted R-squared}: The adjusted R-squared value is 0.81, demonstrating the model's high explanatory power.
\item \textbf{R-squared}: The R-squared value is 0.84, further indicating the model's robustness in fitting the data.
\end{itemize}
The Lag-Llama model benefits from its neural network architecture, which allows it to process complex patterns in time-series data more effectively. The inclusion of probabilistic forecasting with confidence intervals also provides additional insights into the uncertainty of predictions, which is valuable for economic forecasting. From Table \ref{tab:lag_llama_vs_benchmark}, we can see that all the Lag-Llama metrics are lower than 95\%(or higher than 105\% where appropriate) from results of our Benchmark model. 
\begin{table}[htbp]
\centering
\caption{Comparison of Lag-Llama to the Benchmark}
\label{tab:lag_llama_vs_benchmark}
\begin{tabular}{lcc}
\toprule
\textbf{Metric} & \textbf{Lag-Llama} & \textbf{Benchmark} \\
\midrule
MAE & \textbf{0.21} & 0.26 \\
MAPE & \textbf{0.40} & 0.54 \\
MSE & \textbf{0.07} & 0.13 \\
Correlation & \textbf{0.92} & 0.86 \\
R-squared & \textbf{0.84} & 0.73 \\
Adjusted R-squared & \textbf{0.81} & 0.68 \\
\bottomrule
\end{tabular}
\end{table}

\section{Discussion}

Our research question was whether the pre-trained Lag-Llama model can outperform MIDAS regression in the task of inflation nowcasting and how both models compare to a simple benchmark model using previous monthly values of inflation. Firstly, it is important to stress that all three models have been evaluated over a 2-year period between June 2019 and June 2021. The predictions of the MIDAS regression are shown in Figure~\ref{fig:MIDAS_prediction}. The predictions of the Lag-Llama LSTM model are shown in Figure~\ref{fig:Llama_prediction}, together with the confidence intervals as this is a probabilistic time-series prediction.\\

Before starting the comparative analysis, it is important to note that we have given MIDAS the best-case scenario. As a result, the following choices were made. First of all, MIDAS was evaluated only on in-sample predictions and not on out-of-sample predictions, as this approach typically leads to slightly better prediction performance. Another advantage MIDAS received is that MIDAS regression is designed for a specialized task, namely accommodating mixed-frequency data. Lag-Llama, on the other hand, is not designed for mixed-frequency data inputs and as a result, we decided to down-sample our data in order for the Lag-Llama model to still capture some of the information in daily data. The last advantage of MIDAS regression is the inclusion follows from the choice of data variables. Despite the data choice being based on the previous literature, two of our variables turned out to be largely insignificant. This is not a problem for the MIDAS regression as insignificant variables are omitted from the predictions. However, this is not the case for Lag-Llama. Inclusion of data with a low level of information will result in worse predictions for the Lag-Llama model.\\ 

The comparative performance of the MIDAS, Lag-Llama, and benchmark models, as summarized in Table~\ref{tab:model_comparison}, shows overall slightly better performance of Lag-Llama. Despite evaluating the MIDAS regression under optimal conditions, the Lag-Llama model still managed to achieve overall better performance metrics, although the advantage was not uniformly better by the previously defined 5\% margin. One might argue that this definition, where every performance metric needs to be lower than 95\% (or higher than 105\% where appropriate) compared to the MIDAS regression is rather strict. That is indeed the case. However, this research is on a specific dataset with a specific variable selection, therefore more research is needed in order to compare the models. The first performance metric is the MAE. It's lowest for the Lag-Llama, at 0.21 compared to 0.23 for MIDAS and 0.26 for the benchmark model, indicating a more accurate prediction on average. Similarly, the MAPE for Lag-Llama is 0.40, which is lower than the 0.50 observed for MIDAS and 0.54 for the benchmark model, suggesting that Lag-Llama offers a tighter approximation of the actual values in percentage terms. This trend continues with the MSE, where Lag-Llama achieves a lower score of 0.07 compared to 0.11 by MIDAS and 0.13 by the benchmark model, highlighting its greater effectiveness in minimizing the squared differences between predicted and actual values. Furthermore, the correlation coefficient, measuring the linear correspondence between the predicted and actual values, is higher for Lag-Llama at 0.92 compared to 0.88 for MIDAS and 0.73 for the benchmark model. This higher correlation coefficient for Lag-Llama suggests a stronger alignment with the observed data trends.\\

\begin{table}[htbp]
\centering
\caption{Comparison of Lag-Llama to MIDAS Regression}
\label{tab:model_comparison}
\begin{tabular}{lcc}
\toprule
\textbf{Metric} & \textbf{Lag-Llama} & \textbf{MIDAS} \\
\midrule
MAE & \textbf{0.21} & 0.23 \\
MAPE & \textbf{0.40\%} & 0.50\% \\
MSE & \textbf{0.07} & 0.11 \\
Correlation & 0.92 & 0.88 \\
R-squared & \textbf{0.84} & 0.77 \\
Adjusted R-squared & \textbf{0.81} & 0.72 \\
\bottomrule
\end{tabular}
\end{table}
However, comparing with Table \ref{tab:regression_results}, large values for the \textit{USD/EUR exchange rate} (\(\textit{coef} = 2.224\)), \textit{Eurostoxx50\_index} (\(\textit{coef} = -2.087\)), and \textit{EUR/CNY exchange rate} (\(\textit{coef} = -3.621\)) coefficients can be observed. There is a number of reasons that could explain these relatively large numbers. Figure \ref{fig:corr_matrix} shows a slightly higher correlation, 0.49, between \textit{USD/EUR exchange rate} and \textit{EUR/CNY exchange rate}. However, this value is still acceptable according to the standards in the field and does not suggest multicollinearity. The problem with multicollinearity typically starts at correlation levels of around 0.8 among the independent variables \citep{shrestha2020detecting}. Our VIF coefficients in Table \ref{tab:VIF_factors} tell a similar story, as problems with multicollinearity typically emerge at values of 5 or more \citep{kim2019multicollinearity}. Another potential problem might be parameter proliferation. As we have previously mentioned, we are using 22 lags for each independent variable. With seven independent variables, however, only four of them turned out to be significant at the 5\% significance level. This means that we needed to estimate 88 parameters in total. A rule of thumb in machine learning to avoid the curse of dimensionality (also called parameter proliferation in econometrics) is at least 10 observations for each parameter. With our daily data over 11 years, we should have enough parameters to avoid this problem. \\

Another result of our MIDAS regression is that exogenous variables were insignificant by a large margin. Both 3-month LIBOR rates (\(P = 0.475\)) representing short-term interest rates and data on 10-year German Bund (\(P = 0.963\)) turned out to be absolutely insignificant. This is contrary to the findings of \cite{monteforte-2012} and \cite{ekstrom2018importance}, although the exact significance of these variables is unknown. The potential reason for this is that the data on the credit spread already contains similar information about market expectations. This is, however, not supported by the correlation coefficients in Figure \ref{fig:corr_matrix}. Other reasons might be that Germany is not as representative for the Eurozone as was our assumption or that information about short-term and long-term interest rates is not as crucial for inflation nowcasting. Another insignificant variable at the 5\% significance level was our USD/EUR exchange rates (\(P = 0.088\)). One reason for this is that some of the changes in import/export prices were already captured by EUR/CNY exchange rates. One possible improvement in this case is the exclusion of the weekends in our dataset. For the variables where the data is available only for business days, values were filled with linear interpolation. This method, however,  does not add any new information to the regression. As mentioned previously, while this does not influence the performance of MIDAS regression, it has a negative impact on the performance of the Lag-Llama model.\\ 

One notable limitation of using LSTM models in practical applications, particularly within public institutions, is their nature as "black boxes" \citep{rudin2019stop}. The fact that we do not know the reason why an LSTM decides to behave in a certain way can cause problems, such as cases of people wrongly denied parole based on a machine-learning model \citep{wexler2017computer} or machine-learning pollution models stating that dangerous situations are safe \citep{mcgough2018bad}. In regression analysis, each coefficient represents the impact of a specific predictor variable on the dependent variable, allowing for straightforward interpretation and validation against economic theory or prior research. Furthermore, significance levels show us which variables have a significant effect on our dependent variable and which do not. This transparency is crucial for policymakers and analysts who need to justify and communicate their modeling choices and results. On the other hand, LSTMs, due to their complex architecture involving multiple layers and numerous parameters, do not provide the same level of insight into how individual inputs influence the output. This lack of interpretability can be a significant drawback in contexts where understanding the contribution of each variable is essential for decision-making, policy formulation, or regulatory compliance. There is a field of explainable artificial intelligence that aims at tackling this issue. Consequently, despite their superior predictive performance, the adoption of LSTM models is often limited by the need for models that are not only accurate but also transparent and interpretable.\\


Lastly, we would like to point out the use of vocabulary in this study. Although the fields of econometrics and machine learning have some similarities, which can also be seen in the small field of computational econometrics, there is often different vocabulary used for the same phenomenon. In our study, we are using certain words from these two different disciplines interchangeably. In order to avoid confusion, Table \ref{tab:vocab_comparison} summarizes interchangeable words that are used in this paper.\\

\begin{table}[htbp]
\centering
\caption{Vocabulary Differences between Econometrics and Machine Learning}
\label{tab:vocab_comparison}
\begin{tabular}{ll}
\toprule
\textbf{Econometrics} & \textbf{Machine Learning} \\
\midrule
Dependent (endogenous) variable & Target \\
Independent (exogenous) Variable (Predictor) & Feature \\
Error Term & Residual \\
Parameter Proliferation & Curse of Dimensionality \\
Goodness of Fit & Model Evaluation \\
In-sample Predictions & Training Predictions \\
Out-of-sample Predictions & Test or Validation Predictions \\
\bottomrule
\end{tabular}
\end{table}

\section{Conclusion}
The results show that the Lag-Llama model, based on LSTM neural network techniques, has achieved better performance across almost all chosen metrics, including Mean Absolute Error (MAE), Mean Absolute Percentage Error (MAPE), Mean Squared Error (MSE), correlation coefficient, R-squared, and Adjusted R-squared, compared to the MIDAS regression. The one metric that did not turn out to be under 95\% compared to the MIDAS results was the correlation coefficient between predictions and target data. This is shown in Table \ref{tab:model_comparison} where metrics that are lower than 95\% (or higher than 105\% where it is appropriate) are highlighted in bold. We have previously defined that in order to conclude that Lag-Llama has better performance, all the metrics need to be at least 5\% better than the MIDAS regression. Therefore, we conclude that the Lag-Llama model does not outperform the MIDAS regression. Furthermore, under our definition, the MIDAS regression has not been able to outperform our Benchmark model either, as correlation coefficients were within the 5\% range. One conclusion we can draw is that the Lag-Llama model was able to outperform our benchmark model. However, we can see that the integration of advanced machine learning techniques can potentially improve accuracy and timeliness of predictions in economic forecasting. The performance of the Lag-Llama model highlights the potential of using foundation models and machine learning architectures in economic forecasting, which traditionally relies on econometric models.\\

\section*{Acknowledgements}
I would like to thank Oksana Kavatsyuk for critical feedback on this paper. Furthermore, I am grateful for the feedback from students  Jonathan Opitz, Tristan Timpers, Twan Tromp and Nina Kubicová.

\section*{Disclosure statement}
No potential conflict of interest was reported by the author(s).




\bibliography{sample}
\newpage
\appendix
\begin{landscape}
\section{Appendix}

\begin{table}[h]
\centering
\begin{tabular}{lrrrrrrrrr}
\hline
 & \textbf{count} & \textbf{mean} & \textbf{std} & \textbf{min} & \textbf{25\%} & \textbf{50\%} & \textbf{75\%} & \textbf{max} \\
\hline
\textbf{OIL} & 3988.000000 & 76.102565 & 27.087402 & 19.330000 & 53.292500 & 67.825000 & 106.930000 & 126.650000 \\
\textbf{LIBOR\_3M} & 2771.000000 & 0.009151 & 0.318214 & -3.714000 & -0.027467 & 0.005762 & 0.043488 & 10.986014 \\
\textbf{INTEREST\_SPREAD} & 3988.000000 & 0.731351 & 0.546117 & -0.437364 & 0.323852 & 0.693396 & 1.079581 & 2.222250 \\
\textbf{USD/EUR exchange rate} & 2763.000000 & 0.828077 & 0.073159 & 0.674400 & 0.759200 & 0.843800 & 0.892500 & 0.962700 \\
\textbf{Eurostoxx50\_index} & 2534.000000 & 3153.874929 & 440.893087 & 1995.010000 & 2913.420000 & 3220.080000 & 3478.747500 & 4158.140000 \\
\textbf{CNY/EUR exchange rate} & 2528.000000 & 0.126901 & 0.009542 & 0.103800 & 0.120900 & 0.127500 & 0.133000 & 0.152200 \\
\textbf{German\_bund} & 2951.000000 & 0.788072 & 0.990065 & -0.854000 & 0.085000 & 0.503000 & 1.544000 & 3.497000 \\
\textbf{HICP} & 131.000000 & 1.251145 & 0.928967 & -0.600000 & 0.450000 & 1.300000 & 2.000000 & 3.000000 \\
\hline
\end{tabular}
\caption{Statistical Summary of Financial Variables and HICP}
\label{table:stats_summary}
\end{table}

\end{landscape}

\end{document}